# Practical Entropy-Compressed Rank/Select Dictionary

Daisuke Okanohara[*]    Kunihiko Sadakane[†]

**Abstract**

Rank/Select dictionaries are data structures for an ordered set $S \subset \{0, 1, \ldots, n-1\}$ to compute **rank**$(x, S)$ (the number of elements in $S$ which are no greater than $x$), and **select**$(i, S)$ (the $i$-th smallest element in $S$), which are the fundamental components of *succinct data structures* of strings, trees, graphs, etc. In those data structures, however, only asymptotic behavior has been considered and their performance for real data is not satisfactory. In this paper, we propose novel four Rank/Select dictionaries, **esp**, **recrank**, **vcode** and **sdarray**, each of which is small if the number of elements in $S$ is small, and indeed close to $nH_0(S)$ ($H_0(S) \leq 1$ is the zero-th order *empirical entropy* of $S$) in practice, and its query time is superior to the previous ones. Experimental results reveal the characteristics of our data structures and also show that these data structures are superior to existing implementations in both size and query time.

## 1 Introduction

Rank/Select dictionaries are data structures for an ordered set $S \subset \{0, 1, \ldots, n-1\}$ to support the following queries:

- **rank**$(x, S)$: the number of elements in $S$ which are no greater than $x$,
- **select**$(i, S)$: the position of $i$-th smallest element in $S$.

These data structures are used in succinct representations of several data structures. A succinct representation is a method to represent an object from an universe with cardinality $L$ by $(1 + o(1)) \lg L$ bits[1]. While this idea is very similar to the idea of data compression, the difference is that succinct representations support fast queries on the object such as enumerations or navigations. Various succinct representation techniques have been developed to represent data structures such as ordered sets [25, 19, 20, 21], ordinal trees [1, 26, 5, 6, 13, 18, 23], strings [4, 9, 10, 22, 23, 24], functions [17], and labeled trees [1, 3]. All these data structures are based on a succinct representation of Rank/Select dictionaries.

Many data structures have been proposed for Rank/Select dictionaries, most of which support the queries in constant time on word RAM [7, 13, 16, 19, 21] using $n + o(n)$ bits or $nH_0(S) + o(n)$ bits ($H_0(S) \leq 1$ is the zero-th order *empirical entropy* of $S$). In most of these data structures, however, their asymptotic behavior is only considered, and their performance is not optimal for real-size data. As a result, the query time is slow and the data structure size is large for real data. Although recently some practical implementation of Rank/Select dictionaries have been proposed using $n + o(n)$ bits [8, 14], there is no practical implementation of those using $nH_0(S) + o(n)$ bits. Recently *gap-based* compressed dictionaries have been proposed [11, 12]. They use another measure called $gap(S) := \sum_{i=1...m} \lceil \lg(\mathbf{select}(i+1, S) - \mathbf{select}(i, S)) \rceil$ to define the minimum space to store $S$ and propose the data structure using $gap + O(m \log(n/m)/\log m) + O(n \log \log m/n)$ bits, which is much smaller than the entropy-based ones if $m \ll n$, but it cannot not support constant time rank and select queries because of the lower bound [15, 7].

---

[*]Department of Computer Science, University of Tokyo. Hongo 7-3-1, Bunkyo-ku, Tokyo 113-0013, Japan. (hillbig@is.s.u-tokyo.ac.jp).

[†]Department of Computer Science and Communication Engineering, Kyushu University. Motooka 744, Nishi-ku, Fukuoka 819-0395, Japan. (sada@csce.kyushu-u.ac.jp). Work supported in part by the Grant-in-Aid of the Ministry of Education, Science, Sports and Culture of Japan.

[1]Let $\lg n$ denote $\log_2 n$

We will introduce novel four Rank/Select dictionaries, **esp**, **recrank**, **vcode** and **sdarray**(**sarray** and **darray**), each of which is based on different ideas and thus has different advantage and disadvantage in terms of speed, size and simpleness. These sizes are small if the number of elements in $S$ is small, and even close to the zero-th order *empirical entropy* of $S$, $H_0(S) \leq 1$, which is defined as $nH_0(S) = m \lg \frac{n}{m} + (n-m) \lg \frac{n}{n-m}$ where $m$ is the number of elements in $S$.

Table 1 summarizes the properties of proposed data structures for an ordered set $S \subset \{0, 1, \ldots, n-1\}$ with $m$ elements in terms of size, time for **rank** and **select**. We note that these bounds are in the worst case and we can expect faster in practice. For example, the $O(\log^4 m/\log n)$ term in **sarray** and **darray** and $O(\log n)$ term in **vcode** are $O(1)$ in almost the case.

Table 1: The space and time results for **esp**, **recrank**, **vcode**, **sarray** and **darray** for an ordered set $S \subset \{0, 1, \ldots, n-1\}$ with $m$ elements. $H_0(S) \leq 1$ is the zero-th order *empirical entropy* of $S$.

| method | size (bits) | **rank** | **select** |
|---:|:---:|:---:|:---:|
| **esp** (Sec. 3) | $nH_0(S) + o(n)$ | $O(1)$ | $O(1)$ |
| **recrank** (Sec. 4) | $1.44m \lg \frac{n}{m} + m + o(n)$ | $O(\log \frac{n}{m})$ | $O(\log \frac{n}{m})$ |
| **vcode** (Sec. 5) | $m \lg(n/\lg^2 n) + o(n)$ | $O(\log^2 n)$ | $O(\log n)$ |
| **sarray** (Sec. 6) | $m \lg \frac{n}{m} + 2m + o(m)$ | $O(\log \frac{n}{m}) + O(\log^4 m/\log n)$ | $O(\log^4 m/\log n)$ |
| **darray** (Sec. 6) | $n + o(n)$ | $O(1)$ | $O(\log^4 m/\log n)$ |

We conducted experiments using proposed methods and previous methods and show that our data structures are fast and small compared to the previous ones.

## 2 Preliminaries

In this paper we assume the word RAM model. Under the word RAM model we can perform logical and arithmetic operations for two $O(\log n)$-bit integers in constant time, and we can also read/write consecutive $O(\log n)$ bits of memory for any address in constant time.

An ordered set $S$, which is a subset of the universe $U = \{0, 1, \ldots, n-1\}$, can be represented by a bit-vector $B[0, \ldots, n-1]$ such that $B[i] = 1$ if $i \in S$ and $B[i] = 0$ otherwise. We denote $m$ as the number of ones in $B$. Then $\mathbf{rank}(x, S)$ is the number of ones in $B[0, x]$, and $\mathbf{select}(i, S)$ is the position of the $i$-th one from the left in $B$. These values are computed in constant time on word RAM using $O(n \log \log n/\log n)$-bit auxiliary data structures [16].

The above representation of $S$ using the bit vector of length $n$-bit is the worst-case optimal because there exist $2^n$ different sets in the universe and we need $\lg 2^n = n$ bits to distinguish different subsets. We call this representation *verbatim representation*. Similarly, a lower-bound of the size of the representation of $S$ with $m$ elements is $\mathcal{B}(n, m) = \lceil \lg \binom{n}{m} \rceil$ bits. This value is approximately $nH_0(B)$, which is further approximated by $H_0(B) \leq m \lg \frac{n}{m} + 1.44m$ bits. Therefore the size of the *verbatim representation* is far from this lower-bound if $m \ll n$. Raman et al. [21] proposed a constant-time Rank/Select data structure whose size is $\mathcal{B}(n, m) + O(n \log \log n/\log n)$, which matches the above lower-bound asymptotically.

The applications of Rank/Select dictionaries can be divided into two groups. One is for sets with $m \simeq n/2$ and the other is for sets with $m \ll n$. In this paper we call the former *dense sets* and the latter *sparse sets*. Typical applications of *dense sets* are for the wavelet trees [9] that are used for indexing strings, and for ordinal trees. On the other hand *sparse sets* are used in many succinct data structures in order to compress pointers to blocks, each of which stores a part of the data. Because in the word RAM model any consecutive $O(\log n)$ bits of data are accessed in constant time, we often divide the data into blocks of $\Theta(\log n)$ bits each. For example, an ordinal tree with $n$ nodes is encoded in a bit-vector of length $2n$, and to support tree navigating operations, the bit-vector is divided into block of length $\frac{1}{2}\lg n$ bits and in each block we logically mark one bit to construct a contracted tree with $O(n/\log n)$ nodes. These logical marks are represented by a bit-vector of length $2n$ in which $4n/\lg n$ bits are one. The ratio of one is $2/\lg n$, that is, the vector is sparse. Such vectors can be encoded in $\mathcal{B}(2n, 4n/\log n) + O(n \log \log n / \log n) = O(n \log \log n / \log n) = o(n)$ bits. Therefore for storing a sparse vector in a compressed form is important for succinct data structures.

In this paper we will mainly focus on *sparse sets* to support **rank** and **select** functions. Although in some applications like wavelet trees we also need a **select**$_0$ function[2], we usually assume *dense sets* in such applications and well-developed Rank/Select dictionaries for *dense sets* can be applied.

### 2.1 Previous Implementation of Rank/Select Dictionaries

We first give a brief description of Rank/Select dictionary using $n + o(n)$ bits, which is called **verbative**. We conceptually partition $B$ into subsequences of length $l := \log^2 n$ each, called *large block*. Then each *large block* is partitioned into subsequences of length $s := \log n/2$ each, called *small block*. For the boundaries of *large blocks* we store rank-directory (results of **rank**) in $R_l[0 \ldots n/l]$ explicitly using $O(n/\log^2 n \cdot \log n) = O(n/\log n)$ bits. We also store rank-directory for each boundary of small blocks in $R_s[0 \ldots n/s]$, but here we store only relative values to ones stored for the large blocks, which are stored in $O(n \log \log n / \log n)$ bits.

Then **rank** is computed by $\mathbf{rank}(x, S) = R_l[\lfloor x/l \rfloor] + R_s[\lfloor x/s \rfloor] + popcount(\lfloor x/s \rfloor \cdot s, x \bmod s)$, where $popcount(i, j)$ is the number of ones between $B[i \ldots i+j]$ which can be calculated in constant time using a pre-computed table of size $O(\sqrt{n} \log^2 n)$ bits or the *popcount* function [8][3]. For **select** we have two options; the first is a constant time solution using $o(n)$ auxiliary data structures [14] and the second is a $O(\log n)$ solution which is a binary search using **rank** functions without any auxiliary data structures [8]. Because of the luck of space we omit the detail of **select** in constant time [14].

We next introduce Rank/Select dictionary using $nH_0(S) + o(n)$ bits, which is called **ent**. The main difference between **verbative** and **ent** is the representation of bit-vector itself, that is each small block is encoded by the *enumerative code* [2] as follows. Given $t$, the length of the block, and $u$, the number of ones in the block, we calculate $\sum_{i=1 \ldots u} \binom{i}{t-p_i-1}$ where $p_u$ is the position of $i$-th one in the block. This value is the unique number in $[0, \lceil \lg \binom{t}{m} \rceil - 1]$ for each possible block of $t$ length with $u$ ones. This number can be represented by $\mathcal{B}(t, u) = \lceil \lg \binom{t}{m} \rceil$ bits and the size of all encoded blocks is less than $\mathcal{B}(n, m) \leq nH_0(S)$ [19]. We represent each *small block* as the result of *enumerative code*, and the total size is less than $nH_0(S)$. Since they have different sizes, we also need to store pointers to compressed small blocks, which is $O(n \log \log n / \log n) = o(n)$ bits. These encoding and decoding are performed by using pre-computed table of $O(\sqrt{n} \log^2 n)$-bits.

We note that although the size of **ent** is $nH_0(S) + o(n)$ bits, we cannot ignore the $o(n)$ term because $nH_0(S)$ term is small compared to $n$ if $m \ll n$ and o($n$) is as much as $\Theta(nH_0(S))$.

---
[2]We do not discuss **rank**$_0$ since it can be computed by **rank** as $\mathbf{rank}_0(i, S) = i + 1 - \mathbf{rank}(i, S)$.
[3]In this paper let $a \bmod b$ denote $a - \lfloor a/b \rfloor$.

## 3 Estimating Pointer Information

We first propose **esp** (stands for EStimating Pointer information), which does not require *pointer information* by estimating them from **rank** information. Although the size of *pointer information* is O($n \log \log n / \log n$) = o($n$), this size is actually large as much as $\Theta(nH_0(S))$ terms for real-size data.

First we show the propositions which are needed to bound the size of compressed bit vector in terms of **rank** information. Given a bit-vector $B[0 \ldots n-1]$ with $m$ ones, let $L(B)$ be the length of code word for $B$ using *enumerative code* [2] (See Section 2.1). Then,

**Proposition 1** $L(B) \leq H_0(B)$.

Because $H_0(B)$ is the size of a representation of block that uses $\lg(n/m)$ bits for each 1's and $\lg(n/(n-m))$ bits for each 0's, and the $L(B) = \mathcal{B}(n,m) := \lceil \lg \binom{n}{m} \rceil$ is the smallest length of the code to represent the bit vector.

Let $B_i$ ($i = 1 \ldots \lceil \frac{n}{u} \rceil$) be the partition of $B$, and $u$ be the size of each block. Then,

**Proposition 2** $\sum_{i=1}^{\lceil \frac{n}{u} \rceil} L(B_i) \leq \sum_{i=1}^{\lceil \frac{n}{u} \rceil} uH_0(B_i) \leq nH_0(B)$.

The second inequality holds because $nH_0(B)$ is the concave function.

Let $B' := B[0 \ldots t]$ ($t \leq n$) be the prefix of bit-vector $B$. Since $L(B') \leq H_0(B')$ (use Prop.(1) and Prop.(2)), we can store all code words of $B'$ within $H_0(B')$ bits. However since the inequality not equality holds we still have an estimation error of pointers. We therefore need to insert gap bits so that we always estimate the correct pointer information.

We will explain the details of **esp**. Basically, **esp** is based on **ent** except the existence of *super-large blocks* (SLB) since we need to reset estimation errors in each *SLB*. We conceptually partition $B$ into subsequences of length $k := \log^3 n$ each, called *super large block* (SLB). Then each SLB is partitioned into *large block* (LB) of length $l := \log^2 n$. Then each LB is partitioned again into *small block* (SB) of length $s := \log n/2$. We then encode each SB by *enumerative code* (Section 2.1) independently. The code word for $i$-th SB: $SB_i$ is stored in the position which is determined as follows. Let $l_r$ and $s_r$ be results of **rank** for LB and SB as $l_r = R_l[x_l]$, and $s_r = R_s[x_s]$ where $x_l = \lfloor x/l \rfloor$ and $x_s = \lfloor x/s \rfloor$. Then we estimate the starting positions of LB and SB as

$$lp = H_0(LB'_{x_l}) = l_r \cdot \lg \frac{l \cdot x_l}{l_r} + (l \cdot x_l - l_r) \cdot \lg \frac{l \cdot x_l}{l \cdot x_l - l_r} \quad (1)$$

$$sp = H_0(SB'_{x_s}) = s_r \cdot \lg \frac{s \cdot x_s}{s_r} + (s \cdot x_s - s_r) \cdot \lg \frac{s \cdot x_s}{s \cdot x_s - s_r}. \quad (2)$$

where $LB'_{x_l}$ denotes the preceding LBs from the boundary of SLB up to $LB_i$ and $SB'_i$ denotes the preceding SBs from the boundary of LB up to $SB_i$. Then the position for compressed $SB_i$ is $slp + lp + sp$ where $slp$ is the pointer information of SLB which is stored explicitly. We note that all code words are not overlapped (use Prop.(2)) and gap-bits are automatically inserted.

We store rank-directory for LB, SB and *pointer information* for *SLB*. All of them are stored in o($n$) bits.

For **rank**($x, S$), we lookup correspondent rank-directory for LB, SB as $l_r = R_l[x_l]$, and $s_r = R_s[x_s]$ where $x_l = \lfloor x/l \rfloor$ and $x_s = \lfloor x/s \rfloor$. Then we estimate the *pointer information* for LB and SB as in (1) and (2). We then read the compressed bit representation of SB from that position and decode it in constant time and do *popcount* as in **verbative**.

For **select**, we use the same approach as in [14] which is done in constant time with the o($n$)-bits auxiliary data structures.

In practice, since it is very slow to compute the logarithm of a floating-point number for the estimating the entropy, we use a pre-computed table lookup and also use fixed-point integer representation. We require two integer multipliers and one integer addition for estimating one value of the entropy.

## 4 RecRank

The second data structure **recrank** uses the reduction of a sparse bit-array into a contracted bit-array and a *denser* extracted bit-array which was originally used for Algorithm I in [14]. Here we use the reduction recursively.

Given a bit-arrays $B[0 \ldots n-1]$ with $m$ ones, we conceptually partition $B$ into the blocks $B_0, \ldots, B_{n/t}$ of length $t$. We call zero block (ZB) a block where all elements are 0 and non-zero block (NZ) a block where there is at least one 1. The *contracted* bit-array of $B$, $B_c[0, \ldots, n/t-1]$ is defined as a bit-string such that $B_c[i] = 0$ if $B_i$ is ZB, and $B_c[i] = 1$ if $B_i$ is NZ, and the *extracted* bit-array $B_e$ is defined as a bit-array which is formed by concatenating all NZ blocks of $B$ in order.

We can calculate **rank** of $B$ using $B_c$ and $B_e$ as

$$\mathbf{rank}(x, B) = \mathbf{rank}(\mathbf{rank}(\lfloor x/t \rfloor, B_c) \cdot t + (x \bmod t) \cdot B_c[\lfloor x/t \rfloor], B_e). \tag{3}$$

We then recursively apply this reduction by considering the *extracted* bit array as a new input bit array. We continue this process until the *extracted* bit-array is dense enough (the probability of one in a bit-array is larger than $1/4$). After $u$ times of the reduction, we have $t$ contracted bit arrays $B_c^1, B_c^2, .., B_c^t$ and the final *extracted* bit array $B_e^t$.

Here we take the strategy that *contracted* bit arrays would be *dense* (the probability of ones in the bit array would be $1/2$). Let $p(B) = m/n$ be the probability of ones in the bit array $B$. We choose the block size $t = \frac{1}{-\lg(1-p)}$ so that the $p(B_c)$ would be 0.5. This is because the probability of $t$ bits being all zero is $(1-p)^t$ and the half of the elements in *contracted* bit array is one when $(1-p)^t = 1/2$. Then the length of $B_c$ is $-n \lg(1-p)$ and the length of $B_e$ is $n/2$. We note that $B_e$ contains $m$ ones and $p(B_e) = 2p$. This reduction is applied $u = -\lg p$ times so that the probability of ones in the final *extracted* bit array is larger than $1/4$.

Let $T$ be the size of **recrank** and $p = 2^{-u}$. We can calculate $T$ as follows,

$$T = n \cdot \sum_{i=0\ldots u-2} \left( -\frac{\lg(1-2^i p)}{2^i} \right) + 2m \tag{4}$$

$$\leq n \cdot \frac{1}{\log_e(2)} \sum_{i=0\ldots u-2} \left( \frac{2^i p + 2 \cdot (2^i p)^2 / 3}{2^i} \right) + 2m \tag{5}$$

$$= \frac{1}{\log_e(2)} (-m \lg p - 2m/3 - 2mp/3) + 2m \tag{6}$$

In (5), we use $\lg(1-x) \leq x + \frac{2}{3}x^2$ for $0 \leq x \leq \frac{1}{4}$. In short, $T$ is bounded by $1.44 m \lg n/m + m$ bits.

For **rank**, we apply (3) at each stage. Since the number of reduction is $-\lg p = \lg n/m$ and each stage is done in constant time, the total time is $O(\log n/m)$. For **select**, we apply **select** in each stage, each of which is done in constant time [14], so the total time is in $O(\log n/m)$.

```
int select_vc(int i){ // return select(i,S)
    int b = i/p; int q = i%p; // b is the block number and q is the offset
    int x = S[b] + q;
    for (int j = 0; j < T[b]; j++) // count the number of ones in first q bits in each digit
        x += popcount[V[b][j] & ((1U << q) - 1)] << j;
    return x;
}
```

Figure 1: An example code of **select** in **vcode** written in C++. $V[p][j]$ contains $V_p[j]$ and $popcount[k]$ returns the number of set bit in binary sequence of $k$. Other variables correspond to the definition in the paper

## 5 Vertical Code

A *Vertical Code* (**vcode**) supports fast **select** and small space-size in practice because of its byte-based operations and a novel orientation of data. This is a kind of opportunistic data structure, that is, although it is not entropy-compressed Rank/Select dictionary in the worst case, in most case its size is close to the zero-th order *empirical entropy*.

Given a bit-arrays $B[0\ldots n-1]$ with $m$ ones, we first convert it into the *gap* sequence $d[0\ldots m-1]$, $d[i] = \textbf{select}(B, i+1) - \textbf{select}(B, i) - 1$, $(d[0] = \textbf{select}(B, 1))$, $(i = 0\ldots m-1)$.

We then partition $d$ into blocks $B_1, \ldots, B_{m/p}$ of size $p = \mathrm{O}(\log^2 n)$. Let $T[0\ldots m/p-1]$ be the arrays such that $T[i] = \lg\lfloor \max_{j=0\ldots p-1} d[ip+j] \rfloor$, $V_i[j]$ be the bit arrays of length $p$ consisting of the set of the $j$-th bit of $d$ in the block $B_i$, and $S[0\ldots m/p-1]$ be the arrays such that $S[i] = d[ip]$. We note that all $d$ in a block $B_i$ can be represented in $T[i]$ bits each.

We describe how to get **select**$(S, i)$ by using $T$, $V$ and $S$. Let $b = i/p$ and $q = i \bmod p$. Since **select**$(S, i) = S[b] + q + \sum_{i=bp}^{bp+q} d[i]$, we count the number of ones in the first $q$ bits of each $V_b[0], \ldots, V_b[T_i]$, then sums them up with shift. Figure 5 shows the example code of **select** in **vcode**.

The characteristic of **vcode** is if we set $p$ is a multiple of eight, all operations are byte-aligned. And the cost of $\sum_{i=bp}^{bp+q} d[i]$ is $\mathrm{O}(T[b])$, which would be small if $T[b]$ is small. This idea is similar to *gap-based* compressed dictionary [11, 12]. We encode *gap* information directly and we can expect the time of **select** is small if *gap* is small. For example, the *gap* of $\psi$ in compressed suffix arrays [24] is very small.

For **select**, we need to do $T[i]$ operations each of which is done in constant time. Since $T[i]$ would become $\mathrm{O}(\log n)$ in the worst case, the total time for **select** is $\mathrm{O}(\log n)$. For **rank** and **select**$_0$, we need to do the binary search from $m$ elements using **select** which is done in $\mathrm{O}(\log n \cdot \log m)$ time in the worst case.

The size of $S$ is $\mathrm{O}(\log n \cdot m/\log^2 n) = \mathrm{o}(n)$. Since $d[i] < n$, the size of $T_i$ is bounded by $\lg n$ and $T$ is bounded by $\mathrm{O}(\log n \cdot m/\log^2 n) = \mathrm{o}(n)$ and the size of $V$ is bounded by $m \lg n / \lg^2 n$ bits, which happens when $d[ip] = n/\lg^2 n$ $(0 \leq i < n/p)$ and others $d[i]$ are all 0. we note that we can expect the size of $V$ is close to $m \lg n/m$ $(\simeq n H_0(B))$ bits and the time of **select** is close to $\mathrm{O}(1)$ when adjacent elements in $d$ have similar values.

## 6 SDarrays

The idea of SDarrays (**sdarray**) is to use two different techniques for *sparse sets* and *dense sets* each, which enables us to design the data structure simply. We call the former **sarray** and the latter **darray**(**sarray** uses **darray** as a part of data structure).

```
int rank_sarray(int i){ // return rank(i,B) in sarray
    int y = select_0(i/2^w,H)+1; int x = y-i/2^w;
    for (int j = i%2^w; H[y] == 1; x++,y++){
        if (L[x] >= j){ //L is lower-bit of B
            if (L[x] = j) x++;
            break;
        }
    }
    return x;
}
```

Figure 2: An example code of **rank** in **sarray**. Variables correspond to the definition in the paper

First we will introduce **sarray** for *sparse sets*. Given a bit-arrays $B[0\ldots n-1]$ with $m$ ones ($m \ll n$), we define $x[0\ldots m-1]$ such that $x[i] = \textbf{select}(i+1, B)$. Each $x$ is then divided into upper $z = \lfloor \lg m \rfloor$ bits and lower $w = \lceil \lg n/m \rceil$ bits. Lower bits are stored explicitly in $L[0\ldots m-1]$ using $m \cdot \lceil \lg n/m \rceil$ bits. Upper bits are represented by a bit array $H[0\ldots 2m-1]$ such that $H[x_i/2^w + i] = 1$ and others are 0. By using $H$ and $L$, we can calculate **select** in **sarray** by $\textbf{select}(i, B) = (\textbf{select}(i, H) - i) \cdot 2^w + L[i]$. We need **select** for $H$. Here we can assume that $H$ is dense because there are $m$ ones and $m$ zeros in $H$.

We then explain **darray** for *dense sets*, $B[0\ldots n-1]$ with $m \simeq n/2$ ones[4]. We first partition $H$ into the blocks such that each block contains $L$ ones respectively. Let $P_l[0\ldots n/L-1]$ be the bit arrays such that $P_l[i]$ is the position of $(iL+1)$-th one. We classify these blocks into two groups. If the length of block size $(P_l[i] - P_l[i-1])$ is larger than $L_2$, we store all the positions of ones explicitly in $S_l$. If the length of block size is smaller than $L_2$, we store the each $L_3$-th positions of ones in $S_s$. We can store these values in $\lg L_2$ bits.

For $\textbf{select}(i, B)$ in **darray**, we lookup $P_l[\lceil i/L \rceil]$ and see whether the block is larger than $L_2$ or not. If it is, we lookup the value in $S_l$ which is stored explicitly. If not, we lookup correspondent $L_3$-th value in $S_s$ and then do sequential search in the block which would take $O(L_2/\log n)$ time because we can read $O(\log n)$ bits in RAM model. We note that if we can assume that ones is distributed in $B$ uniformly, this sequential search is done in $O(1)$ time. Although this data structure concerns only **select**, we can use same data for $\textbf{select}_0$ by reversing bits in $H$ at reading time.

For **rank** in **darray**, we use the same method as in **verbative**. For $\textbf{rank}(i, B)$ in **sarray**(see the example code in figure 6), we first calculate $y = \textbf{select}_0(i/2^w, H) + 1$ to find the smallest element which is greater than $\lceil i/2^w \rceil \cdot 2^w$. Then we count the number of elements which equals to or smaller than $i$ by sequentially searching over $H$ and $L$ in time $O(n/m)$ because the possible bit pattern of length $\lg n/m$ is $n/m$. If we use binary search, we can do it in $O(\log n/m)$ time but this is slower than sequential search in practice and we use sequential search.

The size of $P_l$ is $O(\frac{n}{L} \cdot \log n)$, that of $S_l$ is at most $\frac{n}{L_2} \cdot L \lg n$ bits, and that of $S_s$ is at most $\frac{n}{L_3} \lg L_2$ bits. When we choose $L := O(\log^2 n)$, $L_2 := O(\log^4 n)$, and $L_3 := O(\lg n)$, all the sizes of $P_l$ and $S_l$ and $S_s$ are $o(n)$. In summary, the size of **darray** is $n + o(n)$ bits.

We then analyze the size of **sarray**. We use $m \cdot \lceil \lg n/m \rceil$ bits for $L$. For $H$ of length $2m$, we use the data structure of **darray**, which is $m + o(m)$ bits. Therefore the total size of **sarray** is $m \cdot \lceil \lg n/m \rceil + 2m + o(m)$ bits.

---

[4]The size of $H$ in **sarray** is $2m$ not $n$. Here we explain **darray** in general case.

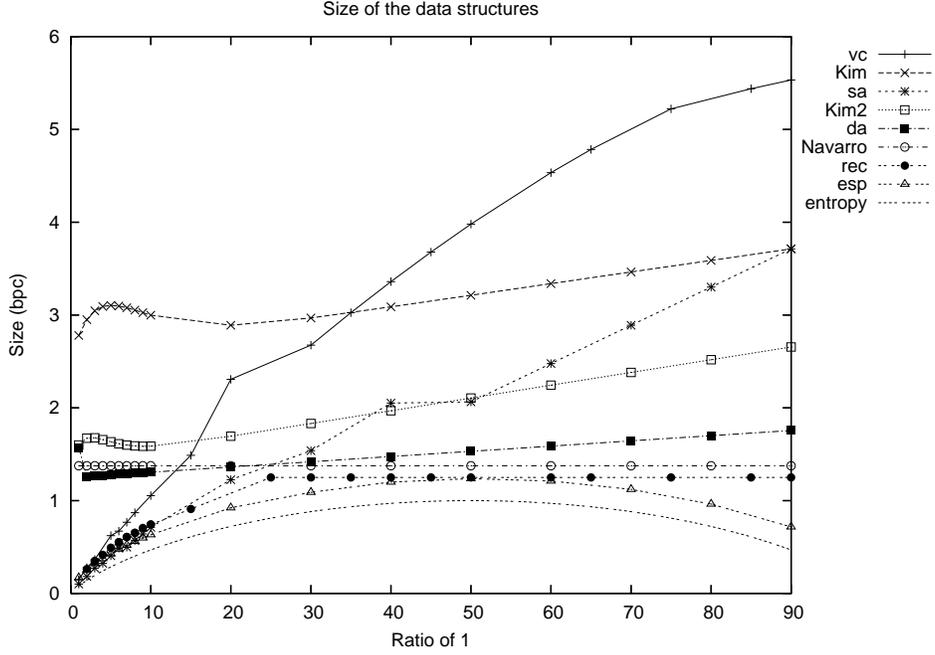

Figure 3: Size of the data structures.

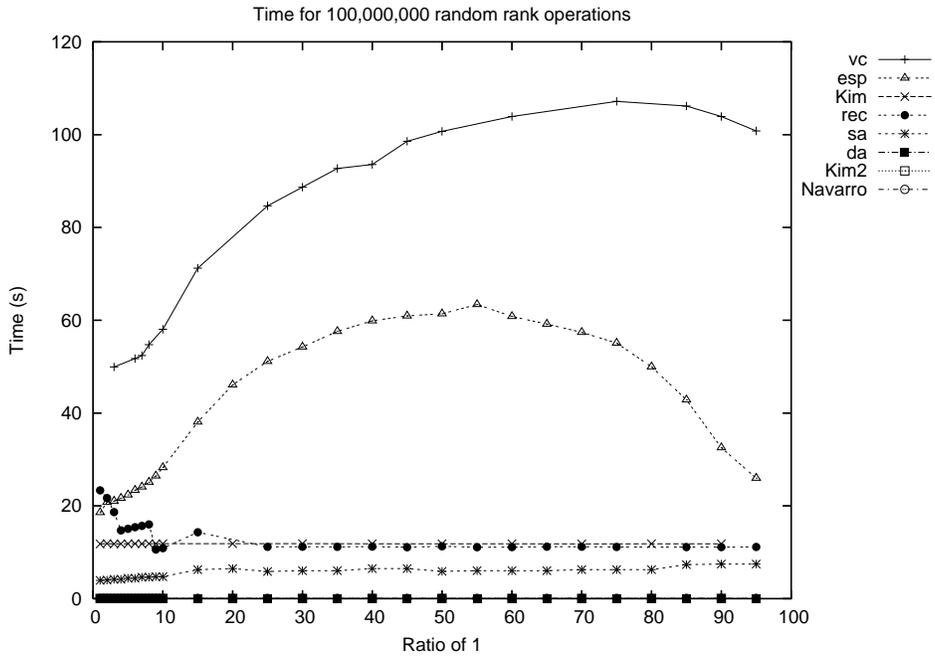

Figure 4: Time for 100,000,000 random rank operations.

## 7 Experimental Results

We conducted experiments using **esp** (*esp*), **recrank** (*rr*), **vcode** (*vc*), **sarray** (*sa*) and **darray** (*da*). We also compare them with byte-based implementation in [14] (*Kim*), and its reimplementation by us (*Kim2*) and [8] (*navarro*). For *esp*, we used $k = 2^{12}$, $l = 2^8$, $s = 2^5$.

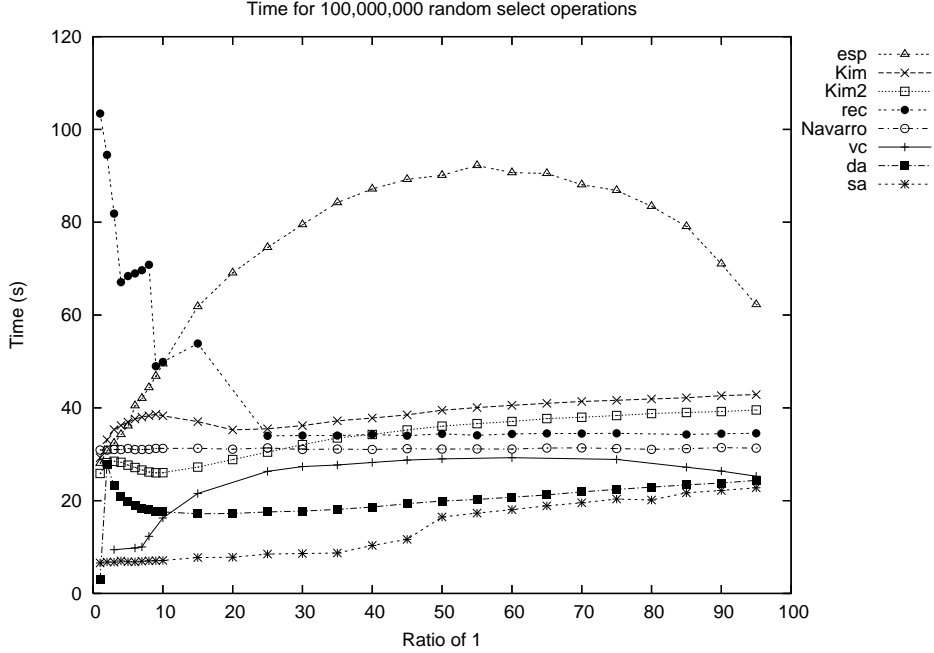

Figure 5: Time for 100,000,000 random select operations.

Table 2: The space results for **esp**, **recrank**, **vcode**, **sarray** and Navarro for the bit arrays of $n$-bit length with 1% and 5% ones. The values is the percentage of the size of each data structures over an original bit-array.

| Ratio of 1's | **esp** | **recrank** | **vcode** | **sarray** | $nH_0$ | Navarro |
|---|---|---|---|---|---|---|
| 1% | 17.02 | 15.83 | 15.05 | 10.13 | 8.08 | 137.5 |
| 5% | 42.67 | 49.32 | 62.25 | 40.59 | 28.64 | 137.5 |

For *vc*, we used $p = 8$. For *sa*, we used $L = 2^{10}$ $L_2 = 2^{16}$ and $L_3 = 2^5$.

For **select** in *rr*, we used $O(\log n)$ solutions because $o(n)$ auxiliary data would become large. For **rank** and **select** in *sa* and *da*, we used sequential search in $H$ because it is faster in practice.

We used GNC C 3.4.3 -O6 -m64. We measured time using *ftime* functions on the 3.4GHz Xeon with 8GB main memory.

All experiments are done using the bit arrays of length $10M(10 \cdot 2^{20})$ bits.

Figure 3 shows the result of the size of several data structure. We also show the result of $nH_0(B)$ which is the lower bound of data structure if we only know the ratio of ones. From here we can see that the size of *esp* is very close to $nH_0(B)$ in all conditions. We also find that the size of *rec*, *sa* and *vc* are very close to $nH_0(B)$ when the ratio of 1 is very small.

Table 2 shows the sizes of each data structures for the bit-arrays with 1% and 5% ones. We find that the sizes of proposed data structures are indeed close to $nH_0$. We note that **sarray** is the smallest in both case.

Figure 4 is the result of $10^8$ **rank** operations. We can see that *Kim2, Navarro* and *da* is the fastest which is the same as in **rank** in **verbative**. On the other hand *vc* is the slowest for **rank** because it needs binary search using **select** functions. Only *rec* is slower in the small ratio of 1 because its computation cost is $O(\log n/m)$ depending on the inverse number of $m$.

Figure 5 is the result of $10^8$ **select** operations. Among several methods, *sa* is the fastest in all conditions. As in the result of **rank**, *rec* is slower in the small ratio of 1. We also find that *da* shows different behavior in the small ratio of 1 because it switches data structures depends on the ratio of 1. We note that the result of *esp* for **rank** and **select** is fast in the ratio of 1 is small or large sinc *esp* employ decode table for *enumerative code* which is only prepared for compressible block. Therefore it becomes slower when the block could not be compressible.

We did not show the results in bit arrays of different length because of the lack of space. We note that except *Navarro*, *rec* and *vc* which use binary search in **select**, all methods have the similar result with bit arrays of different length.

## 8 Concluding Remarks

In this paper, we have proposed novel four Rank/Select dictionaries, **esp**, **recrank**, **vcode** and **sdarray**. Experimental results show that the sizes of these data structures are indeed close to the zero-th order *empirical entropy* and achieves fast queries.

We also note that they are easy to implement (except **esp**) because **recrank** uses reduction which can employ well-developed *dense sets* techniques and **vcode** converts the problem into the *popcount* in bytes and **sdarray** separates the problem for *dense sets* and *sparse sets*, which simplify the problem.

In the next stage of our research, we will extend our result to more complex data structures, such as sequences from large alphabets. We also consider applications which employ appropriate data structures and also apply them to data compression as well.

## References

[1] D. Benoit, E. D. Demaine, J. I. Munro, R. Raman, V. Raman, and S. S. Rao. Representing trees of higher degree. *Algorithmica*, 43(4):275–292, 2005.

[2] T. Cover. Enumerative source encoding. *IEEE Trans. on Information*, 19(1):73–77, 1973.

[3] P. Ferragina, F. Luccio, G. Manzini, and S. Muthukrishnan. Structuring labeled trees for optimal succinctness, and beyond. In *FOCS*, 2005.

[4] P. Ferragina and G. manzini. Indexing compressed texts. *Journal of the ACM*, 52(4):552–581, 2005.

[5] R. Geary., N. Rahman., R. Raman., and V. Raman. A simple optimal represengtation for balanced parentheses. In *Proc. of CPM*, pages 159–172, 2004.

[6] R. Geary., N. Rahman., and V. Raman. Succinct ordinal trees with level-ancestor queries. In *ACM-SIAM SODA*, pages 1–10, 2004.

[7] A. Golynski. Optimal lower bounds for rank and select indexes. In *Proc. of ICALP*, 2006.

[8] Rodrigo González, Szymon Grabowski, Veli Mäkinen, and Gonzalo Navarro. Practical implementation of rank and select queries. In *Poster Proceedings Volume of 4th Workshop on Efficient and Experimental Algorithms (WEA'05)*, pages 27–38, Greece, 2005. CTI Press and Ellinika Grammata.

[9] R. Grossi., A. Gupta., and J. Vitter. High-order entropy-compressed text indexes. In *Proc. of SODA*, pages 841–850, 2003.

[10] R. Grossi., A. Gupta., and J. Vitter. When indexing equals compression: Experiments with compressing suffix arrays and applications. In *Proc. of SODA*, pages 636–645, 2004.

[11] A. Gupta, W. Hon, R. Shar, and J. Vitter. Compressed data structures: Dictionaries and data-aware measures. In *Proc. of DCC*, pages 213–222. IEEE, 2006.

[12] A. Gupta, W. Hon, R. Shar, and J. Vitter. Compressed dictionaries: Space measures, data sets, and experiments. In *Proc. of WEA*, 2006. To appear.

[13] G. Jacobson. Space-efficient static trees and graphs. In *Proc. of FOCS*, pages 549–554, 1989.

[14] D. K. Kim., J.C. Na., J.E. Kim., and K. Park. Efficient implementation of rank and select functions for succinct representation. In *Proc. of WEA*, 2005.

[15] P. B. Miltersen. Lower bounds on the size of selection and rank indexes. In *Proc. of SODA*, pages 11–12, 2005.

[16] J. I. Munro. Tables. In *Proc. of FSTTCS*, pages 37–42, 1996.

[17] J. I. Munro and S. S. Rao. Succinct representations of functions. In *Proc. of ICALP*, pages 1006–1015, 2004.

[18] J. I. Munro, V. Rman, and S. S. Rao. Space efficient suffix trees. *Journal of Algorithms*, 39(2):205–222, 2001.

[19] R. Pagh. Low redundancy in static dictionaries with constant query time. *SIAM J. Computation*, 31(2):353–363, 2001.

[20] C. K. Poon and W. K. Yiu. Opportunistic data structures for range queries. In *Proc. of COCOON*, pages 560–569, 2005.

[21] R. Raman, V. Raman, and S. S. Rao. Succinct indexable dictionaries with applications to encoding k-ary trees and multisets. In *Proc. of SODA*, pages 232–242, 2002.

[22] S. S. Rao. Time-space trade-offs for compressed suffix arrays. *Information Processing Letters*, 82(6):307–311, 2002.

[23] K. Sadakane. Succinct representations of lcp information and improvements in the compressed suffi arrays. In *ACM-SIAM SODA*, pages 225–232, 2002.

[24] K. Sadakane. New text indexing functionalities of the compressed suffix arrays. *J. Algorithms*, 48(2):294–313, 2003.

[25] K. Sadakane W. K. Hon and W.K. Sung. Succinct data structures for searchable partial sums. In *Proc. of ISAAC*, pages 505–516, 2003.

[26] C. C Lin Y. T. Chiang and H. I. Lu. Orderly spanning trees with applications. *SIAM Journal on Computing*, 34(4):924–945, 2005.